\documentclass[pdflatex,sn-mathphys-num]{sn-jnl}
\usepackage{graphicx}%
\usepackage{multirow}%
\usepackage{amsmath,amssymb,amsfonts}%
\usepackage{amsthm}%
\usepackage{mathrsfs}%
\usepackage[title]{appendix}%
\usepackage{xcolor}%
\usepackage{textcomp}%
\usepackage{manyfoot}%
\usepackage{booktabs}%
\usepackage{algorithm}%
\usepackage{algorithmicx}%
\usepackage{algpseudocode}%
\usepackage{listings}%

\theoremstyle{thmstyleone}%

\theoremstyle{thmstyletwo}%

\theoremstyle{thmstylethree}%

\begin{document}

\title[Article Title]{Physics of Collectivity and EOS from the RHIC Beam Energy Scan Program}

\author*[1,2]{\fnm{Xionghong} \sur{He}}\email{hexh@impcas.ac.cn}

\author*[3]{\fnm{Shusu} \sur{Shi}}\email{shiss@mail.ccnu.edu.cn}

\author*[3]{\fnm{Nu} \sur{Xu}}\email{nuxu02@gmail.com}

\affil[1]{\orgdiv{Institute of Modern Physics}, \orgname{Chinese Academy of Sciences}, \orgaddress{\city{Lanzhou}, \postcode{730000}, \country{China}}}

\affil[2]{\orgdiv{School of Nuclear Science and Technology}, \orgname{University of Chinese Academy of Sciences}, \orgaddress{\city{Beijing}, \postcode{100049}, \country{China}}}

\affil[3]{\orgdiv{Key Laboratory of Quark \& Lepton Physics (MOE) and Institute of Particle Physics}, \orgname{ Central China Normal University}, \orgaddress{\city{Wuhan}, \postcode{430079}, \country{China}}}

\abstract{In this article we will review recent measurements of directed flow $v_1$ and elliptic flow $v_2$
in Au+Au collisions from the STAR Beam Energy Scan (BES) program. We systematically analyze the $v_1$ distributions for identified hadrons ($\pi^\pm$, $K^\pm$, $p/\bar{p}$) and $\Lambda$ hyperon as functions of  rapidity ($y$), with particular focus on the mid-central collisions. The energy dependence of the $v_1$ slope is extracted across the BES range ($\sqrt{s_{NN}}$ = 3 -- 200 GeV). The atomic mass number ($A$) dependence of light and hyper nuclei $v_1$ to test the validity of the coalescence production mechanism.
The constituent quark number (NCQ) scaling is systematically investigated based on $v_2$ measurements of identified particles and strange hadrons. We find that the NCQ scaling approximately holds in Au+Au collisions when $\sqrt{s_{NN}} \geq$  4.5 GeV, but completely breaks down at $\sqrt{s_{NN}}$ = 3.0 and 3.2 GeV. The gradual restoration of NCQ scaling from 3.2 to 4.5 GeV suggests a possible transition in the dominant degrees of freedom from hadrons to partons. 
The physics of collectivity, equation of the system and relevance to the QCD phase diagram will be discussed within the framework of both hydrodynamic and hadronic transport model calculations.}




\maketitle

\section{Introduction}\label{sec1}
The study of hot and dense Quantum Chromodynamics (QCD) matter created in relativistic heavy-ion collisions provides crucial insight into the emergent collective behavior of strongly interacting systems and the underlying equation of state (EOS) of nuclear matter. 
At sufficiently high temperatures and energy densities, QCD predicts that hadronic matter undergoes a smooth crossover transition to a deconfined state of quarks and gluons, known as the Quark-Gluon Plasma (QGP)~\cite{Lee:1974ma, Cabibbo:1975ig}.
Mapping out the QCD phase diagram and locating possible phase transitions and critical phenomena are central goals of the RHIC Beam Energy Scan (BES) program~\cite{Fukushima:2010bq, Bzdak:2019pkr, Luo:2020pef, Chen:2024aom}.

The STAR experiment at RHIC has conducted a systematic energy scan of Au+Au collisions over a broad range of center-of-mass energies ($\sqrt{s_{NN}}$ = 3 -- 62.4~GeV), enabling exploration of the QCD matter under varying temperature and baryon chemical potential. 
One of the key probes of the system's early-stage dynamics and collective evolution is anisotropic flow, characterized by the Fourier coefficients $v_n$ of the final-state azimuthal particle distributions relative to the reaction plane~\cite{Voloshin:2008dg}
\begin{equation}
E\frac{d^{3}N}{dp^{3}}  = 
\frac{1}{2\pi}\frac{d^{2}N}{p_{T}dp_{T}dy}\left( 1 + \sum_{n=1}^{\infty} 2v_n \cos {n(\phi - \Psi)} \right), 
\label{eq:Fourier}
\end{equation}
where $\phi$ is the azimuth of particle and $\Psi$ is the azimuth of the collision reaction plane. 

In particular, the directed flow $v_1$ and elliptic flow $v_2$ carry sensitive information about the equation of state (EOS), including the underlying degrees of freedom, as well as the pressure gradients developed during the collision and the transport properties of the system.

In this article, we present a review of recent directed and elliptic flow measurements from the STAR BES program. The energy dependence of $v_1$ slopes for identified hadrons, including $\pi^\pm$, $K^\pm$, $p/\bar{p}$, and $\Lambda$, is systematically investigated to search for possible softening of the EOS or signatures of a first-order phase transition~\cite{Danielewicz:2002pu, Lattimer:2006xb, Rischke:1995ir, Stoecker:2004qu,STAR:2014clz, STAR:2017okv}. 
Additionally, the mass-number dependence of light nuclei $v_1$ provides an opportunity to test the coalescence mechanism of particle production at intermediate energies.
Elliptic flow measurements for various particle species, including strange hadrons, are used to examine the scaling behavior with the number of constituent quarks (NCQ). 
This scaling, particularly observed in multi-strange hadrons and $\phi$ mesons, is regarded as a hallmark of partonic collectivity in the early stages of the collision~\cite{STAR:2003wqp, STAR:2015gge, STAR:2017kkh}. 
The breakdown and subsequent gradual restoration of NCQ scaling across the BES energy range, most notably between $\sqrt{s_{NN}}$ = 3.0 and 4.5~GeV, may signal a transition in the dominant degrees of freedom from hadronic to partonic matter~\cite{STAR:2025owm}.

Through comparisons with hydrodynamic and hadronic transport model calculations, we aim to gain further understanding of the collective dynamics, the effective EOS, and their implications for the structure of the QCD phase diagram.

\section{Directed Flow $v_1$}
\label{sec:v1}
Directed flow, $v_1$, has two components: a rapidity-even
function $v_1^{\rm even}$ and a rapidity-odd function $v_1^{\rm odd}$.
The $v_1^{\rm even}$ arises from the event-by-event initial nuclei geometry fluctuations~\cite{teaney2011}. Here, we focus only
on the rapidity-odd component which is generated by the hydrodynamic bounce-off of the colliding nuclei (hereafter denoted as $v_1$). 

\subsection{Directed flow of Hadrons}
\begin{figure}
\centering
\includegraphics[width=10 cm]{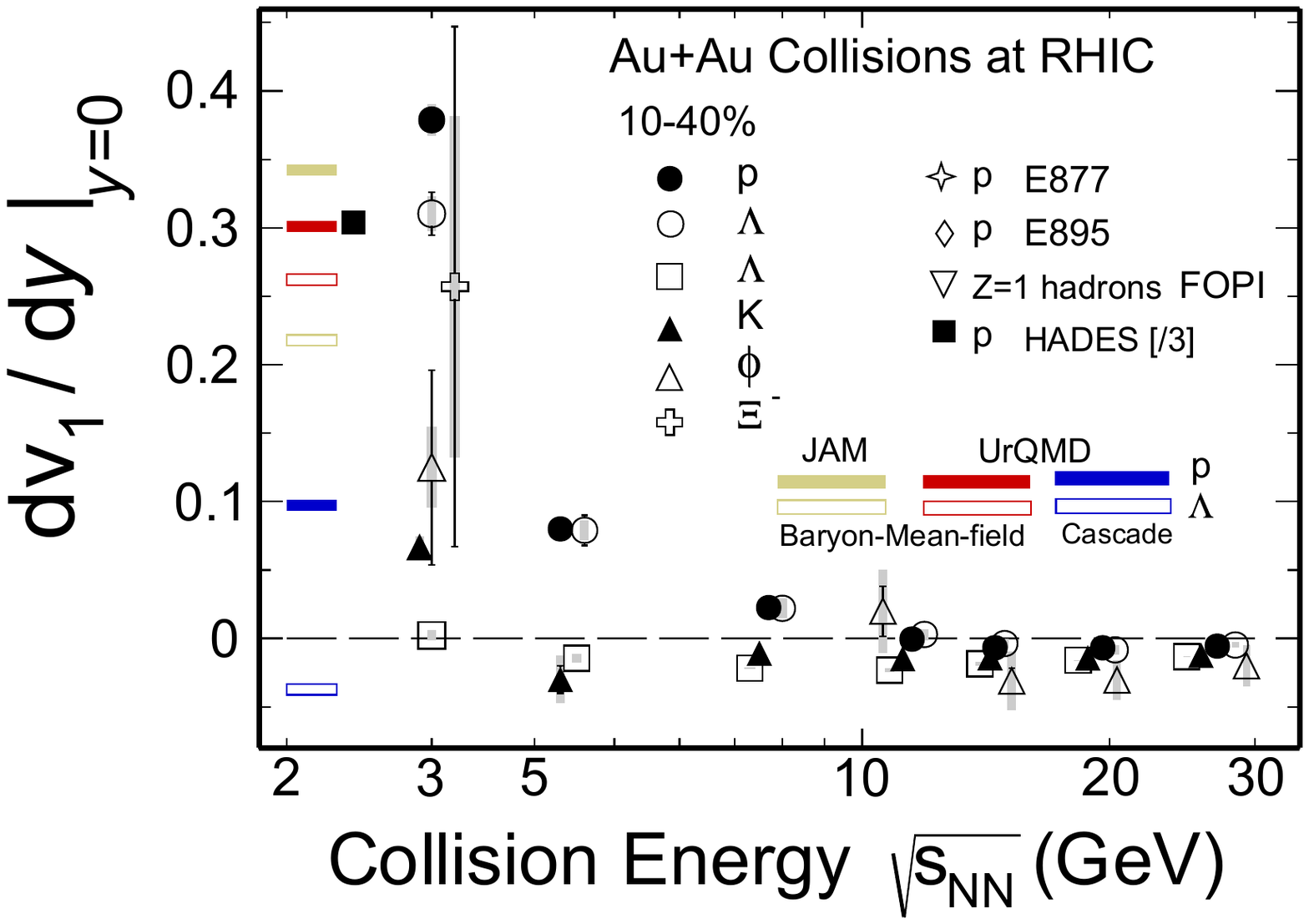}
\caption{Collision energy dependence of directed flow slope $dv_1/dy$ for $\pi$, $K$, proton, $\Lambda$, $\phi$, and $\Xi$ in 10\%–40\% Au+Au collisions(adapted from~\cite{STAR:2021yiu}). Colored bands are the calculations from transport models JAM and UrQMD.
}\label{v1_energy}
\end{figure}

The strength of the directed flow is typically characterized by e $dv_1/dy$, the slope fitted to the $v_1(y)$ $|y| \lesssim 0.5$ to 1.0. Figure~\ref{v1_energy} presents the slope $dv_1/dy$ fitted at midrapidity for 10-40\% centrality Au+Au collisions as a function of beam energy over the range $\sqrt{s_{NN}} =$ 3 -- 27 GeV~\cite{STAR:2021yiu,2025arXiv250323665S}.
At $\sqrt{s_{NN}}=3$ GeV, both the JAM and UrQMD hadronic transport models, when incorporating baryonic mean-field interactions, successfully reproduce the observed trends in the $dv_1/dy$ for protons and $\Lambda$ hyperons.
This consistency between hadronic transport models and experimental data suggests that the dominant degrees of freedom at 
$\sqrt{s_{NN}}=3$ GeV are interacting baryons, which may indicate that partonic degrees of freedom play a negligible role at this collision energy, marking a potential signature of the disappearance of the QGP in the QCD phase diagram.

At collision energies above $\sqrt{s_{NN}} = 11.5$ GeV, the $dv_1/dy$ becomes negative for all measured hadron species. Notably, both protons and $\Lambda$ hyperons exhibit a minimum in $dv_1/dy$ within the $\sqrt{s_{NN}} \sim 10 - 20$ GeV range~\cite{STAR:2014clz,STAR:2017okv}. This  minimum has been theoretically proposed as a potential signature of a first-order phase transition between hadronic matter and QGP~\cite{STOCKER2005121}. Alternative interpretations suggest that hadronic transport models incorporating baryonic mean-field potentials can also qualitatively describe this feature~\cite{RN442}, highlighting the ongoing debate regarding the microscopic origin of this phenomenon in this energy regime.

\begin{figure}
\centering
\includegraphics[width=9 cm]{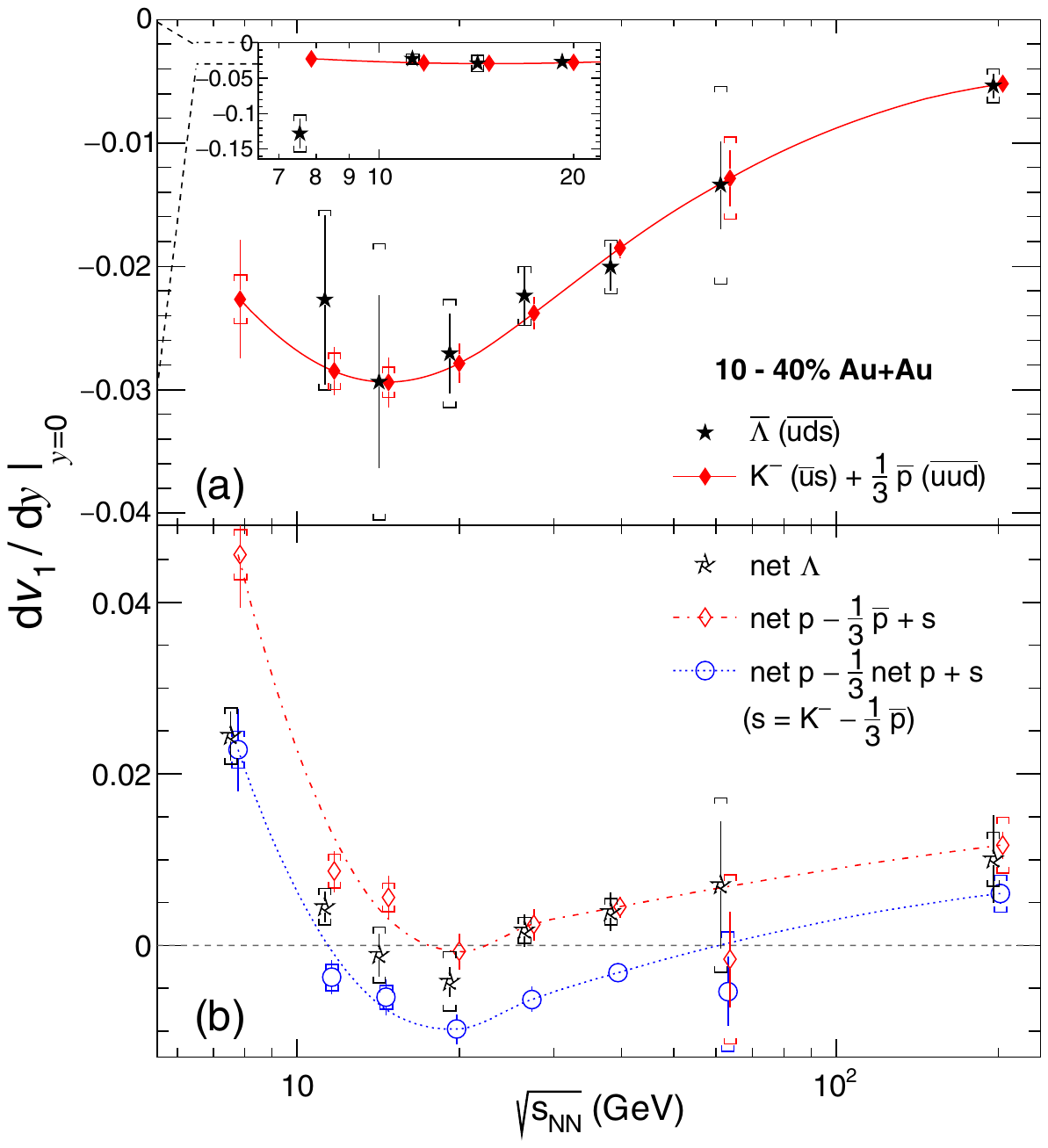}
\caption{Test for coalescence sum rule of directed flow for (a) produced quarks and (b) net-baryon in 10\%–40\% Au+Au collisions(adapted from~\cite{STAR:2017okv}).
}\label{v1_coalescence}
\end{figure}

In the scenario where a QGP phase is formed, the collective flow patterns are initially imprinted on the quark degrees of freedom. As these quarks subsequently coalesce to form hadrons during hadronization, the flow coefficients $v_n$ of the resulting mesons and baryons reflect the sum of the $v_n$ of their constituent quarks. This additive behavior, valid in the limit of small flow coefficients, gives rise to the characteristic number-of-constituent-quark (NCQ) scaling behavior in heavy-ion collisions.

In the collisions, all light antiquarks ($\bar{u}$, $\bar{d}$) and strange quarks ($s$, $\bar{s}$) are produced via pair production, in contrast to $u$ and $d$ quarks, which may either be newly produced or transported from the colliding nuclei. 
To specifically test the NCQ scaling scenario, where all quarks are produced, panel (a) of Fig.~\ref{v1_coalescence} compares $dv_1/dy$ of the anti-$\Lambda$ ($\overline{uds}$) with a weighted combination of $K^-$ ($\bar{u}s$) and $\bar{p}$ ($\overline{uud}$) flows,
$K^-(\bar{u}s) + {1 \over 3}\bar{p}\,(\overline{uud})$,
where the 1/3 weighting factor accounts for the assumption of similar flow patterns for $\bar{u}$ and $\bar{d}$. Strikingly, this combination shows excellent agreement with the anti-$\Lambda$ data across the energy range $\sqrt{s_{NN}} = 11.5-200$ GeV, providing strong evidence for NCQ scaling in the directed flow of produced (anti)quarks.
To enhance the contribution of transported quarks over those produced, net particles are defined as the excess yield of particles over antiparticles.
The panel Fig.~\ref{v1_coalescence} shows the same comparison for $dv_1/dy$ of net-$\Lambda$ defined based on $v_{1\Lambda}=r(y)v_{1\bar{\Lambda}}+[1-r(y)]v_{1{\rm net}\Lambda}$, where $r(y)$ is ratio of observed $\bar{\Lambda}$ to $\Lambda$ yield at each beam energy~\cite{STAR:2017okv}. The sum rule calculation agrees with the net-$\Lambda$ above 11.5 GeV and shows the large discrepancy at 7.7 GeV.

\subsection{Directed flow of Light-/Hypernuclei}
Besides hadrons, heavy-ion collisions also produce light nuclei and hypernuclei in abundance, particularly in the energy range of several GeV, where the production yields of light nuclei (such as $d$, $t$, and $^3$He) become comparable to those of protons~\cite{RN709}. The yield of hypernuclei (e.g., $^3_\Lambda$H, $^4_\Lambda$H) exhibits a maximum around $\sqrt{s_{\rm NN}} = 3-4$ GeV~\cite{RN9}.
These loosely bound objects, consisting of nucleons and hyperons, have binding energies on the order of a few MeV, and their underlying production mechanisms remain a topic of ongoing debate. The study of their formation in heavy-ion collisions provides crucial insights into the properties of nuclear matter under extreme conditions~\cite{RN590, Sun_2021}.

Similar to the NCQ scaling in hadron flow, light-/hyper- nuclei collective flow is theoretically predicted to follow an analogous mass-number (A) scaling under the nucleon coalescence framework.
At $\sqrt{s_{\rm NN}}=$ 3 GeV, the $dv_1/dy|_{y=0}$ follow an approximate $A$ scaling for light nuclei species with $A=1-4$~\cite{RN418}, while above 11.5 GeV, protons develop negative flow whereas deuterons maintain positive values but with larger uncertainties~\cite{RN271}. This breakdown of $A$ scaling between the low-energy ($<$7.7 GeV) and high-energy regimes may suggests either: (1) a transition in light nuclei production mechanisms , or (2) fundamental changes in system evolution dynamics across this energy range.

\begin{figure}
\centering
\includegraphics[width=9 cm]{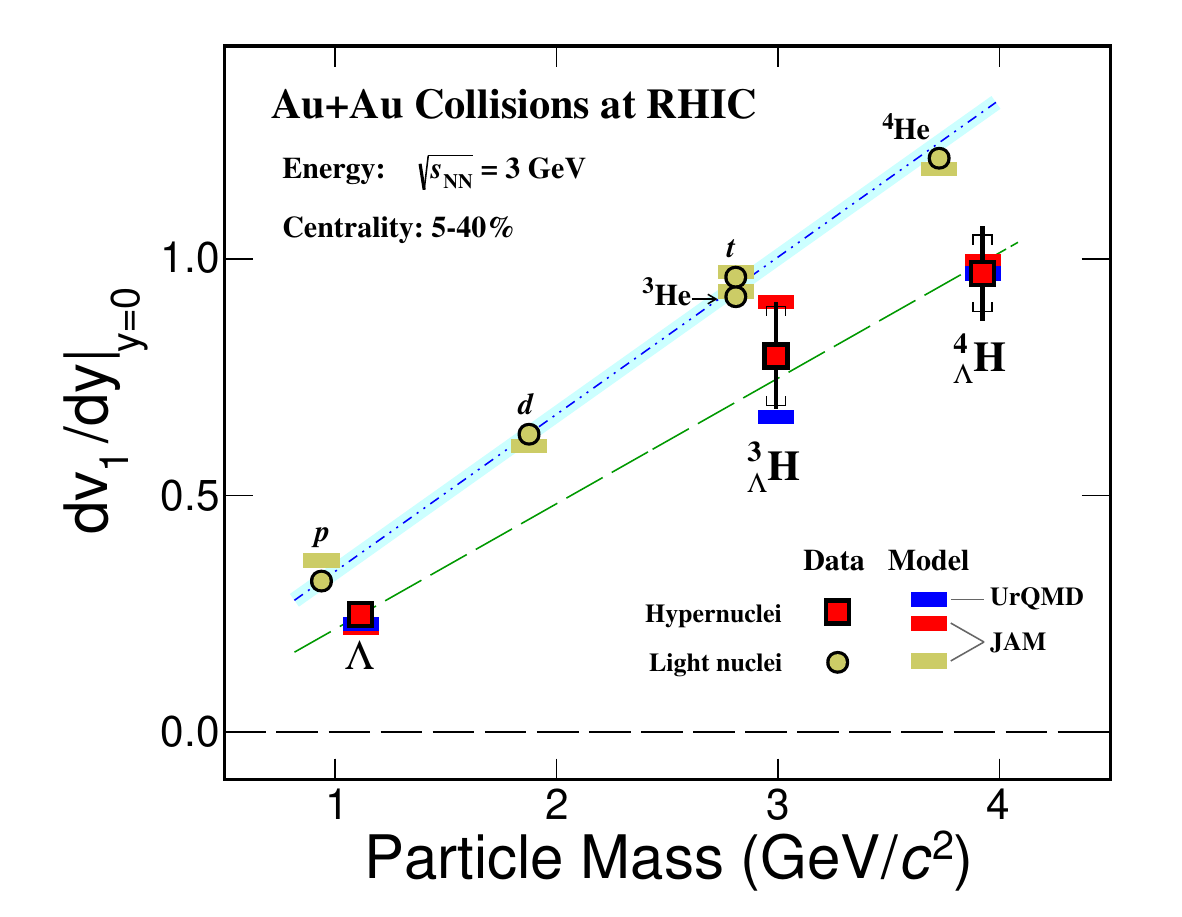}
\caption{Mass dependence of light nuclei and hypernuclei  $dv_1/dy|_{y=0}$  at $\sqrt{s_{\rm NN}}=3$ GeV 5\%-40\% centrality Au + Au collisions(adapted from~\cite{RN510}). The calculations of transport models (JAM and UrQMD) plus coalescence are shown bars.
}\label{v1_hyperNuclei}
\end{figure}

The STAR experiment reported the first observation of the $v_1$ of hypernuclei ${}^3_{\Lambda}$H and ${}^4_{\Lambda}$H in $\sqrt{s_{\rm NN}}=$ 3 GeV Au+Au collisions~\cite{RN510}, as shown in Fig.~\ref{v1_hyperNuclei}.
The directed flow slope ($dv_1/dy|_{y=0}$) of $\Lambda$ hyperons and hypernuclei exhibits a clear mass-number ($A$) scaling analogous to light nuclei, with linearly increasing magnitude proportional to particle mass. Transport model calculations incorporating an afterburner coalescence qualitatively reproduce this scaling behavior, supporting a production mechanism where hypernuclei form via coalescence between hyperons and light nuclear cores during the late stages of heavy-ion collisions. While the scaling slope for hypernuclei differs slightly from that of ordinary nuclei (though current experimental uncertainties remain substantial). This distinction likely originates from differences between nucleon-nucleon ($N$-$N$) and hyperon-nucleon ($Y$-$N$) interaction. Such interactions are particularly crucial for understanding structure of neutron star.

\section{Elliptic Flow $v_2$}
\label{sec:v2}

Elliptic flow ($v_2$) arises from the pressure-driven anisotropic expansion of QCD matter created in non-central heavy-ion collisions. 
It reflects the transport properties and the relevant degrees of freedom of the medium in its early stages.
In particular, the scaling behaviors and particle-type dependencies of $v_2$ serve as sensitive probes of collectivity and the transition from hadronic to partonic matter~\cite{Molnar:2003ff}. 
In this section, we present elliptic flow measurements from four stages of the RHIC program:  
(1) top-energy collisions at $\sqrt{s_{NN}} = 200$~GeV;  
(2) the first phase of the Beam Energy Scan (BES-I), covering $\sqrt{s_{NN}} = 7.7$--62.4~GeV;  
(3) the second phase (BES-II), spanning $\sqrt{s_{NN}} = 3.0$--19.6~GeV; and  
(4) the extraction of the temperature-dependent shear viscosity-to-entropy ratio, $4\pi\eta/s$, from energy-dependent flow data, offering insight into the transport properties and phase structure of the QCD medium.

\subsection{Elliptic Flow at $\sqrt{s_{NN}} = 200$~GeV}

\begin{figure}
\centering
\includegraphics[width=12 cm]{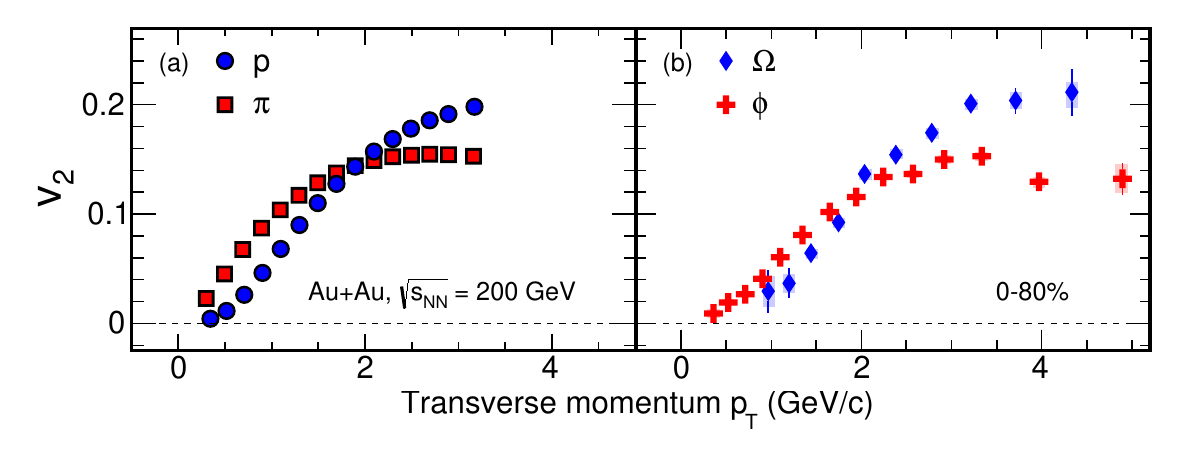}
\caption{$v_2$ as a function of $p_{\rm T}$ for $\pi$ and $p$ (panel a), and for $\phi$ and $\Omega$ (panel b) in minimum-bias Au+Au collisions at $\sqrt{s_{NN}} = 200$~GeV  (adapted from~\cite{STAR:2015gge}).
}\label{phiOmegav2}
\end{figure}

\begin{figure}
\centering
\includegraphics[width=9 cm]{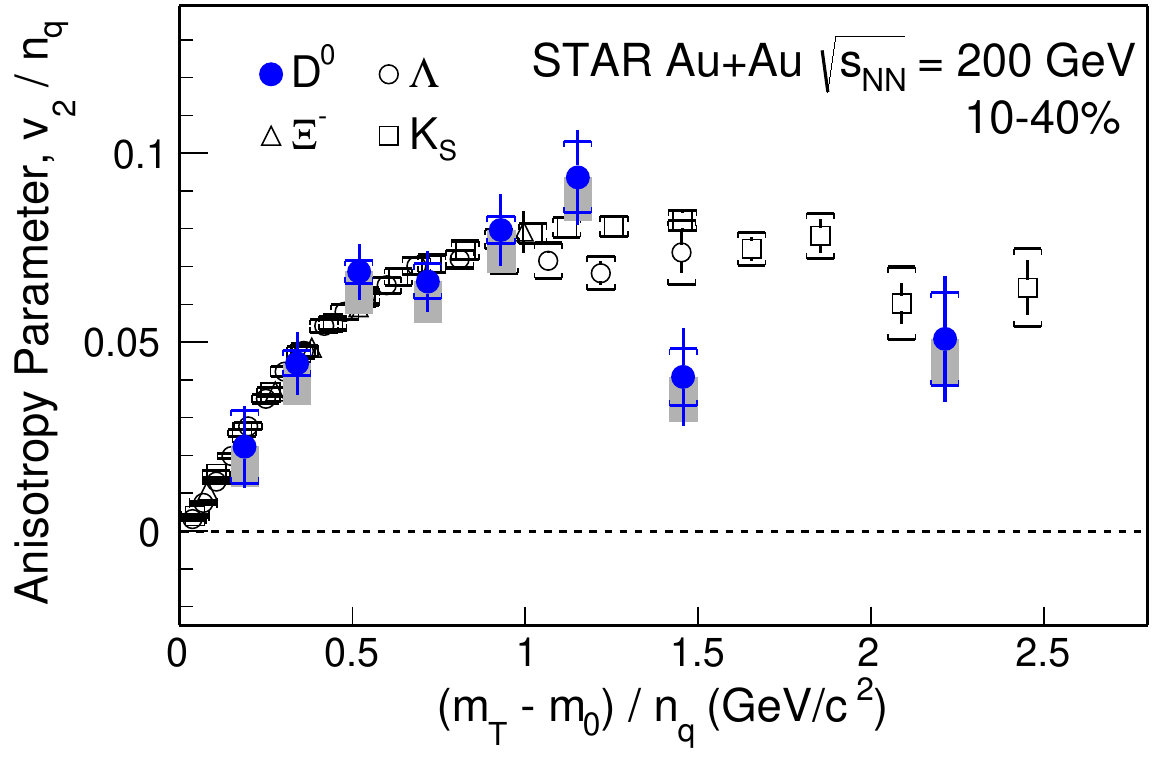}
\caption{The number-of-constituent-quark ($n_q$) scaled $v_2$ as a function of $(m_T - m_0)/n_q$ in 10\%--40\% Au+Au collisions for $D^0$, $\Xi^-$, $\Lambda$, and $K_S^0$ (adapted from~\cite{STAR:2017kkh}).  
Here, $m_T = \sqrt{p_{\rm T}^2 + m_0^2}$ is the transverse mass.
}\label{D0v2}
\end{figure}

At the top RHIC energy of 200~GeV, STAR observed large $v_2$ values for a wide range of identified particles, including light hadrons ($\pi^\pm$, $p/\bar{p}$), strange hadrons ($K^\pm$, $K^0_S$, $\Lambda$), multi-strange hadrons $\Xi$, $\Omega$), the $\phi$ meson and $D^0$ meson~\cite{STAR:2003wqp, STAR:2015gge}. 

These results established a strong signature of collective behavior in the created medium.
A key feature is the observation of mass ordering at low $p_{\rm T}$, consistent with hydrodynamic and transport expansion. 
At intermediate $p_{\rm T}$ ($2 < p_{\rm T} < 5$~GeV/$c$), a baryon-meson splitting is observed, which is successfully described by quark coalescence models. When $v_2$ and $p_{\rm T}$ are scaled by the number of constituent quarks $n_q$, the results for mesons and baryons collapse onto a common curve. This empirical NCQ scaling indicates that the collectivity develops at the partonic level, prior to hadronization.

Figure~\ref{phiOmegav2} illustrates the transverse momentum dependence of $v_2$ for $\pi$, $p$, $\phi$, and $\Omega$ in minimum-bias Au+Au collisions at $\sqrt{s_{NN}} = 200$~GeV. 
The sizable $v_2$ values of $\phi$ and $\Omega$, despite their small hadronic interaction cross sections, highlight the dominance of early partonic collectivity.
In addition, STAR has measured significant $v_2$ for open charm hadrons, specifically the $D^0$ meson, in Au+Au collisions at $\sqrt{s_{NN}} = 200$~GeV~\cite{STAR:2017kkh}. 
Figure~\ref{D0v2} shows the scaling behavior of $v_2$ with the number of constituent quarks for various hadrons, including $D^0$ mesons. 
The $D^0$ meson, containing a charm quark, is produced predominantly in initial hard scatterings and thus experiences the full evolution of the medium. 
The observation of sizable $D^0$ $v_2$ indicates that charm quarks participate in the collective flow, suggesting strong interactions with the medium and partial thermalization of charm.
Furthermore, the approximate NCQ scaling of $D^0$ $v_2$ alongside light and strange hadrons strengthens the interpretation of partonic collectivity and provides important constraints on the heavy-quark diffusion coefficient in the QGP.

\subsection{Elliptic Flow in BES-I: $\sqrt{s_{NN}}$ = 7.7 - 62.4~GeV}

\begin{figure}
\centering
\includegraphics[width=12 cm]{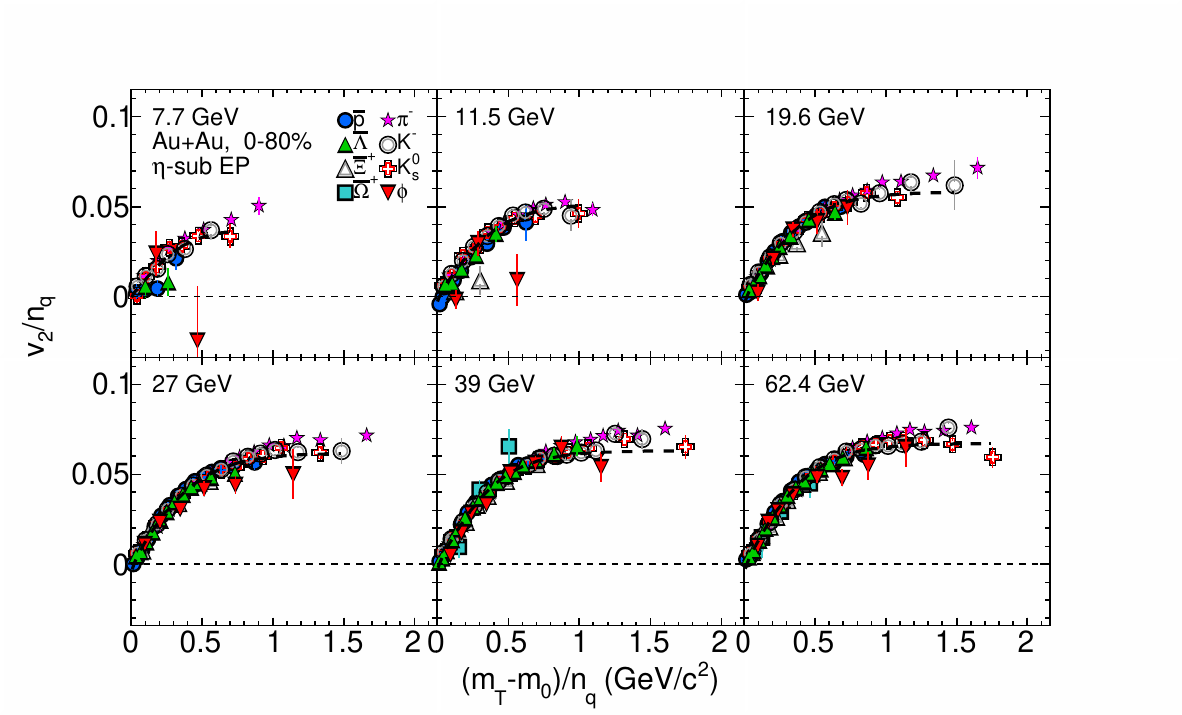}
\caption{Number-of-Constituent-Quark (NCQ) scaled $v_2/n_q$ vs. $(m_T - m_0)/n_q$ for selected particles in 0–80\% central Au+Au collisions at $\sqrt{s_{NN}}$ = 7.7–62.4~GeV (adapted from~\cite{STAR:2013ayu}). 
}\label{BESI-v2NCQ}
\end{figure}

The first phase of the RHIC Beam Energy Scan (BES-I) extended the measurement of $v_2$ to lower energies to search for the onset of deconfinement and possible signatures of a QCD phase transition~\cite{STAR:2013cow, STAR:2013ayu, STAR:2015rxv, STAR:2012och}.  As shown in Fig.~\ref{BESI-v2NCQ}, the NCQ-scaled $v_2$ for various hadrons in Au+Au collisions at $\sqrt{s_{NN}} = 7.7$–62.4 GeV exhibits an approximate scaling behavior across most energies. The measurements of $v_2$ for identified particles from $\sqrt{s_{NN}} = 7.7$ to $62.4$~GeV revealed several important trends.

First, while $v_2$ values decrease gradually with decreasing energy, the qualitative features observed at 200~GeV persist down to at least $14.5$~GeV~\cite{STAR:2015rxv}. 
These include mass ordering, baryon-meson splitting, and approximate NCQ scaling. 
Notably, $\phi$ meson still exhibits non-zero $v_2$ at $\sqrt{s_{NN}} \geq 14.5$~GeV, suggesting that partonic collectivity remains relevant in this energy regime~\cite{STAR:2015rxv}.

At $\sqrt{s_{NN}}$ = 7.7 and 11.5~GeV, the statistical uncertainties are significantly larger, particularly for multi-strange hadrons and $\phi$ mesons, which limits the ability to quantitatively assess the validity of NCQ scaling. 
Within BES-I uncertainties, the scaling behavior remains approximately consistent with expectations, but definitive conclusions about the presence or absence of scaling violations cannot be drawn. 
Nevertheless, comparisons across the BES-I energy range suggest that any potential breakdown of NCQ scaling, if present, would likely become more prominent at energies below $7.7$~GeV. 
These observations motivate high-statistics measurements in the BES-II program to better delineate the transition between hadronic and partonic dominance.

\subsection{Elliptic Flow in BES-II: $\sqrt{s_{NN}}$ = 3.0 - 19.6~GeV}

The BES-II with STAR fixed-target program have extended $v_2$ measurements to even lower energies. 
At $\sqrt{s_{NN}} = 3.0$~GeV, the elliptic flow of identified particles shows qualitatively different behavior compared to higher energies. 
The $v_2$ values for mesons and baryons diverge significantly, and NCQ scaling is completely broken~\cite{STAR:2021yiu}, in stark contrast to its behavior at higher energies. 
This strongly suggests that the medium is dominated by hadronic interactions, with negligible contribution from partonic collectivity.

\begin{figure}
\centering
\includegraphics[width=12 cm]{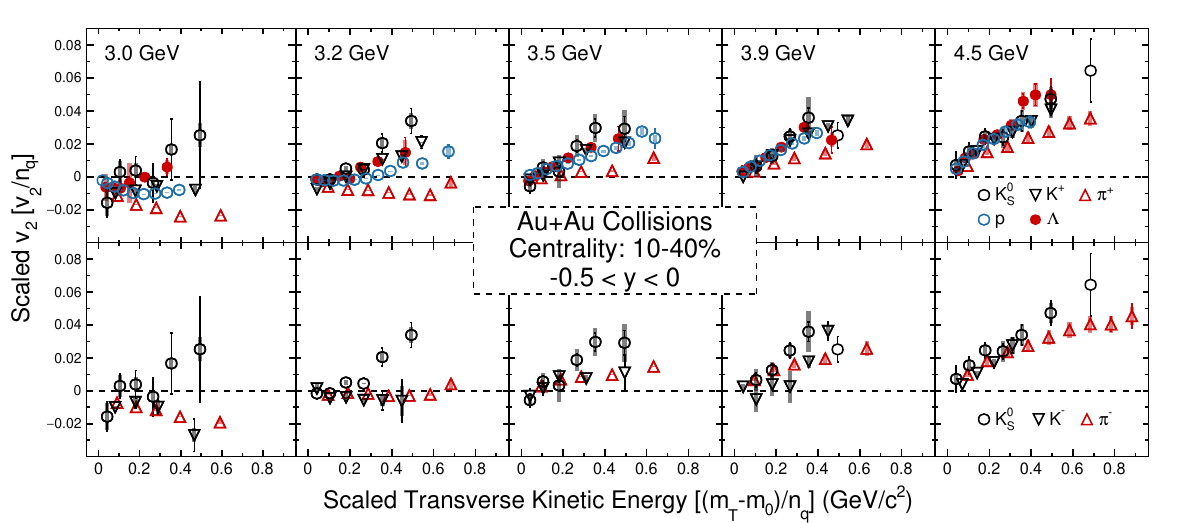}
\caption{NCQ-scaled $v_2$ vs. NCQ-scaled transverse kinetic energy for particles (top) and antiparticles (bottom) in 10–40\% central Au+Au collisions at $\sqrt{s_{NN}} = 3.0$–$4.5$~GeV (adapted from~\cite{STAR:2025owm}). 
}\label{OnsetNCQ}
\end{figure}

Figure~\ref{OnsetNCQ} presents the detailed evolution of NCQ scaling in 10\%–40\% Au+Au collisions across $\sqrt{s_{NN}} = 3.0$–4.5 GeV. 
Between $\sqrt{s_{NN}} = 3.2$ and $4.5$~GeV, a gradual restoration of NCQ scaling is observed~\cite{STAR:2025owm}. 
The $v_2$ values of light and strange hadrons begin to align more closely with the common NCQ-scaled curve. 
This trend implies a transition in the dominant degrees of freedom: from a hadronic phase to a partonic phase. 
Recent preliminary measurements with improved precision for the $\phi$ meson and multi-strange hadrons indicate that NCQ scaling holds in Au+Au collisions at $\sqrt{s_{NN}}$ = 7.7 and 11.5~GeV~\cite{STAR:2025QM}. 
These new results help complete the picture of collectivity and the evolution of dominant degrees of freedom in the QCD medium, covering the full energy range from 200~GeV down to 3~GeV.

These observations place the hadron-parton transition region within $3.2 \lesssim \sqrt{s_{NN}} \lesssim 4.5$~GeV. The measured $v_2$ patterns in this regime provide unique input to theoretical models and help constrain the nature of the QCD phase transition. 
In particular, hydrodynamic models with different equations of state and hadronic transport models will be actively tested against these data to explore the role of the EOS in collective dynamics.

\subsection{Signature of Phase Transition}

\begin{figure}
\centering
\includegraphics[width=10 cm]{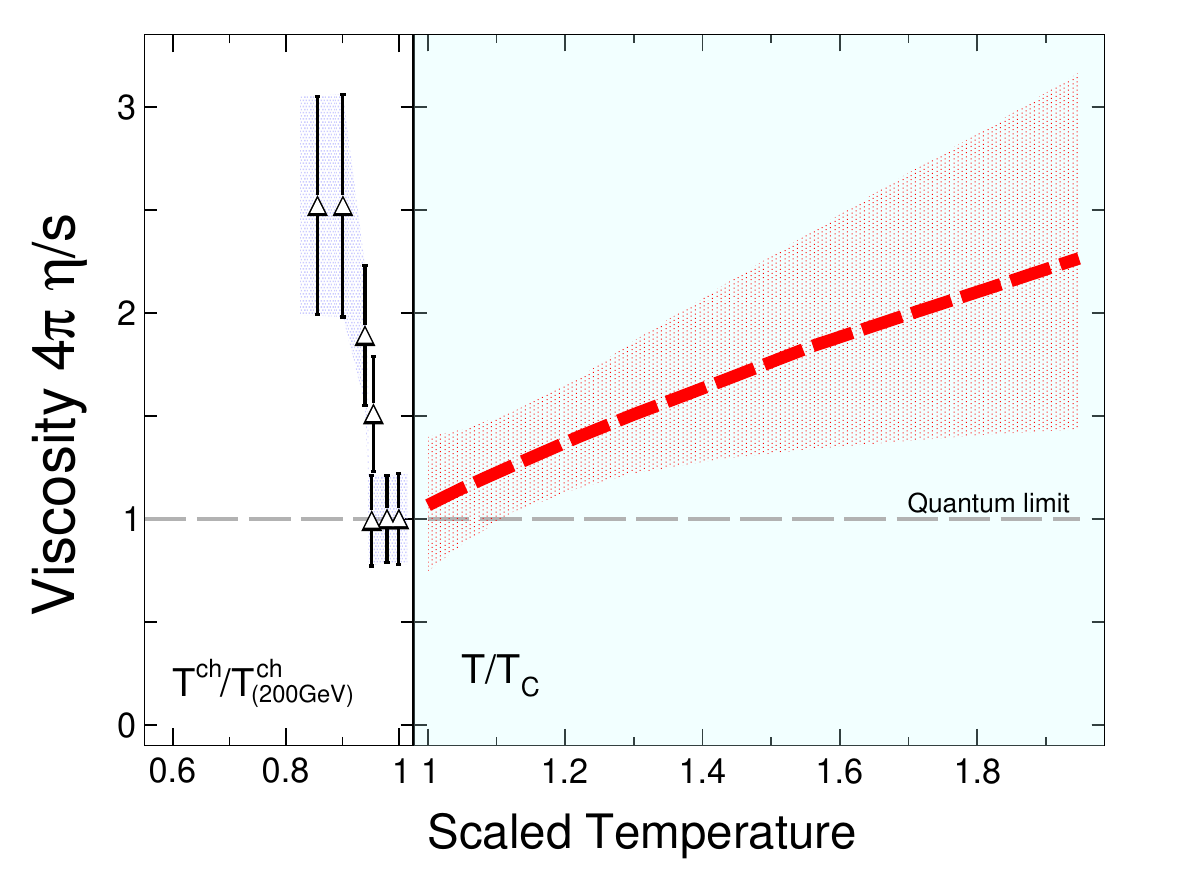}
\caption{Effective shear viscosity-to-entropy ratio ($4\pi \eta/s$) as a function of scaled temperature.
Left: Values extracted from the energy dependence of $v_2$ and $v_3$ using scaled chemical freeze-out temperatures;
Right: Temperature evolution from Bayesian analyses.
The horizontal dashed line marks the quantum lower bound (adapted from~\cite{Huang:2023ibh}).
}\label{viscosity}
\end{figure}

The energy-dependent measurements of collective flow observables offer critical insights into the transport properties and microscopic structure of the strongly interacting medium formed in relativistic heavy-ion collisions. In particular, the anisotropic flow coefficients, such as the elliptic flow $v_2$, provide a sensitive probe for extracting the shear viscosity-to-entropy density ratio $4\pi\eta/s$, which characterizes the medium's dissipative behavior and degree of fluidity.

Figure~\ref{viscosity} illustrates the temperature dependence of the event-averaged $4\pi\eta/s$, as extracted from hydrodynamic model comparisons to flow observables at various beam energies~\cite{Karpenko:2015xea}. In the left panel, the effective values of $4\pi\eta/s$ are shown as a function of the scaled chemical freeze-out temperature, $T_{\mathrm{ch}}/T_{\mathrm{ch}}(200~\mathrm{GeV})$, where $T_{\mathrm{ch}}$ for each energy is obtained from statistical thermal model fits to hadron yields~\cite{STAR:2017sal}. At high collision energies, specifically in the range $\sqrt{s_{NN}}$ = 39 - 200~GeV, the extracted $4\pi\eta/s$ approaches unity, which corresponds to the conjectured quantum lower bound. This observation suggests that the medium created at these energies is dominated by partonic degrees of freedom and exhibits nearly ideal fluid behavior with minimal shear resistance.
In contrast, as the beam energy decreases, hadronic interactions become increasingly significant, leading to a rapid rise in the effective $4\pi\eta/s$. This trend reflects a transition from a strongly-coupled QGP to a more viscous hadronic medium, consistent with expectations for a crossover from deconfined to confined matter.

The right panel of Figure~\ref{viscosity}, reproduced from Ref.\cite{Bernhard:2019bmu}, presents the temperature evolution of $4\pi\eta/s$ as a function of the scaled temperature $T/T_{c}$, where $T_{c}$ represents the critical temperature associated with the QCD crossover transition\cite{Bernhard:2019bmu, Xu:2017obm}. The characteristic $V$-shaped structure, with a minimum near $T/T_{c} \approx 1$, is qualitatively consistent with theoretical expectations from lattice QCD and hydrodynamic modeling~\cite{Csernai:2006zz}. This feature provides experimental support for the existence of a smooth crossover in the QCD phase diagram and highlights the critical region as the point of minimal shear viscosity, where the system behaves as a nearly perfect fluid.

\section{Summary and Outlook}
\label{sec:summary}
In summary, the RHIC Beam Energy Scan (BES) program has provided valuable insights into the properties of strongly interacting matter at high baryon density. 
The observed Number-of-Constituent-Quark (NCQ) scaling of $dv_1/dy$ at $\sqrt{s_{NN}} > 7.7$~GeV indicates that partonic degrees of freedom dominate the QCD medium at higher energies. 
In contrast, at $\sqrt{s_{NN}} = 3$~GeV, the breakdown of this scaling, together with the successful reproduction of $v_1$ by hadronic transport models, demonstrates that hadronic interactions govern the medium at lower energies~\cite{STAR:2021yiu, STAR:2021ozh, Lan:2022rrc}. 
These findings place stringent constraints on the collision dynamics and the QCD equation of state across a wide range of beam energies.

Elliptic flow ($v_2$) has served as a complementary and sensitive probe of collectivity, with systematic studies across quark flavors — from light to multi-strange hadrons ($K^0_S$, $\phi$, $\Lambda$, $\Xi$, $\Omega$) and even charm hadrons — establishing the presence of partonic collectivity at top RHIC energies~\cite{STAR:2015gge, STAR:2017kkh}. 
Multi-strange hadron collectivity has been firmly observed down to $\sqrt{s_{NN}} \approx 39$~GeV~\cite{STAR:2013cow}, providing strong evidence for a strongly interacting QGP. 
At much lower energies, partonic collectivity disappears, as demonstrated by the breakdown of NCQ scaling at $\sqrt{s_{NN}} = 3.0$~GeV~\cite{STAR:2021yiu}. 
Recent STAR measurements show a gradual reemergence of NCQ scaling at $\sqrt{s_{NN}} = 4.5$~GeV~\cite{STAR:2025owm}, suggesting the onset of parton-level collectivity in this regime. 
Collectively, the observed disappearance and subsequent onset of NCQ scaling provide a compelling signal of the transition from hadronic to partonic dominance, completing a coherent narrative of the QGP discovery program at RHIC that spans its discovery, disappearance, and reemergence across different energies.

Overall, $v_1$ and $v_2$ provide complementary constraints on the QCD medium: $v_1$ is highly sensitive to the nuclear equation of state, while $v_2$ and its NCQ scaling behavior serve as key signatures of the emergence of partonic collectivity. This dual perspective offers a comprehensive picture of the QCD phase structure across different energy regimes.
In addition, flow measurements across a broad energy range also provide important information on the transport properties of the medium. 
The extraction of the shear viscosity to entropy density ratio, $4\pi \eta/s$, from comparisons between flow observables and hydrodynamic calculations reveals a characteristic minimum near the QCD critical temperature, consistent with lattice QCD and Bayesian analyses. 
This observation not only demonstrates that the medium created at high collision energies behaves as a nearly perfect fluid dominated by partonic degrees of freedom, but also highlights the rapid increase of $4\pi \eta/s$ toward lower energies where hadronic interactions prevail. 
These findings support the existence of a smooth crossover transition in the QCD phase diagram and establish viscosity as a key dynamical signature of the hadron–parton phase transition.

Future studies motivated by these findings will enable comprehensive investigations of QCD matter at even lower beam energies and in complementary collision systems, where upcoming experimental facilities will play a pivotal role. 
The CEE experiment at HIRFL and HIAF will extend flow measurements to energies below $\sqrt{s_{NN}} = 3$~GeV~\cite{Lu:2016htm, Liu:2023xhc, Zhang:2023hht}, where $v_1$ is expected to reach its maximum, thereby providing decisive information on the hadronic EOS; with its capability to cover a wide range of collision systems from light nuclei (C+C) to heavy nuclei (U+U), CEE will also enable systematic studies of the system-size dependence of collective flow and light-nuclei production. 
In parallel, the NICA-MPD experiment will explore the intermediate energy range ($\sqrt{s_{NN}} \approx 4$-11~GeV) with high luminosity, focusing on collective flow, fluctuations, and multi-strange hadrons~\cite{Kisiel:2020spj, MPD:2025jzd}, thereby probing the dynamics of the hadron-parton transition and possible signatures of a first-order phase transition and critical phenomena in the QCD phase diagram. 
Alongside these efforts, the FAIR-CBM experiment will provide high-precision data on rare probes such as multi-strange hyperons, charm hadrons, and hypernuclei in the $\sqrt{s_{NN}} \approx 2$-8~GeV region~\cite{CBM:2016kpk, CBM:2025voh}; its ability to operate at extremely high interaction rates will yield unprecedented statistical accuracy in flow measurements, offering stringent tests of coalescence models, baryon-hyperon interactions, and the nuclear EOS relevant for neutron star physics.

The BES program and forthcoming experimental initiatives are expected to further advance our understanding of the phase structure of QCD matter, elucidate the nature of the hadron-parton transition, and provide unique insights into the behavior of matter under extreme conditions similar to those in the early universe and the interiors of compact stars.

\section{Acknowledgements}
Shusu Shi is supported in part by the National Key Research and Development Program of China under Grant Nos. 2024YFA1610700 and 2022YFA1604900; and the National Natural Science Foundation of
China under Grant No. 12175084. Xionghong He is supported by the the National Natural Science Foundation of
China under Grant No. 12205342.

\bibliography{ref}%
\end{document}